\documentclass[12pt]{article}

\begin{document}

\title{Noncommutative Planar Particles: Higher Order Versus First
  Order Formalism  and Supersymmetrization}

\author{J. Lukierski\footnote{Talk given by J. Lukierski} 
\\ 
Institute for Theoretical Physics, University of Wroc{\l}aw,
 \\
   pl. Maxa Borna 9, 50-204 Wroc{\l}aw, Poland
  \\
  e-mail:lukier@ift.uni.wroc.pl
\\ \\
P. Stichel
\\
An der Krebskuhle 21, D-33619 Bielefeld, Germany   
 \\
 e-mail:pstichel@gmx.de
 \\ \\
W.J. Zakrzewski
 \\
Department of Mathematical Sciences,
 Science Laboratories,   \\
University of Durham, South Road,
Durham DH1 3LE, UK   
 \\
 e-mail: W.J.Zakrzewski@durham.ac.uk
}

\date{}
\maketitle

\begin{abstract}

We describe the supersymmetrization of two formulations of free
noncommutative planar particles -- in coordinate space with
 higher order Lagrangian [1] and in the framework of
 Faddeev and Jackiw [2,3], with first order action.
 In nonsupersymmetric case the first formulation after imposing 
subsidiary
  condition    eliminating internal degrees of freedom
   provides the second formulation. In supersymmetric case
 one can also introduce the split into ``external'' and ``internal''
  degrees of freedom both describing supersymmetric models.

\end{abstract}


\section{Introduction}

In [1] the present authors introduced the following nonrelativistic
higher order
 action for D=2 (planar) particle:
\begin{equation}\label{pary2.1}
 { L}^{(0)}_1  =
 \frac{m\dot{x}^2_1}{2} - k \epsilon_{ij} \dot{x}_i \ddot{x}_j\, .
 \end{equation}
The canonical quantization of (\ref{pary2.1}) implies the 
consideration of
$x_i, \dot{x}_i$ as independent degrees of freedom (see
e.g. [4]) with the following canonically conjugated   two momenta 
[1,4]

\begin{equation}\label{pary2.2}
p_i = \frac{\partial { L^{(0)}}}{\partial \dot{x}_i} -
\frac{d}{dt}
 \frac{\partial { L^{(0)}}}{\partial \ddot{x}_i}
= m  \dot{x}_i - 2k \epsilon_{ij} \ddot{x}_j \, ,
\end{equation}

\begin{equation}\label{pary2.3}
\widehat{p}_i = k \epsilon_{ij} \dot{x}_j \, .
\end{equation}
The relation (\ref{pary2.3}) introduces a second class constraint, 
i.e. after the
 introduction
 of Dirac brackets
 the
   Lagrangian
   system  (\ref{pary2.1}) is described by six degrees of freedom  
$Y_A =(
x_i, p_i, v_i = \dot{x}_i)$.
   One
 gets the following set of
 Dirac brackets [1]:
 \begin{equation}\label{pary2.4}
 \left\{ Y_A, Y_B \right\} =
 \begin{pmatrix}
 {0 & 1_2 &  0\cr
 -1_2 &0 & 0 \cr
 0 & 0 & - \frac{1}{2k} \epsilon}
 \end{pmatrix}
 \end{equation}

 In order to get the first order formulation of the action (\ref{pary2.1}) 
one can use
 the technique  proposed by Faddeev and Jackiw [2,3]. The action 
(\ref{pary2.1})
  is equivalent to the following one\footnote{The equivalence of (5) 
   and (\ref{pary2.1}) can be seen
   in a clear way if we consider the generating functionals based
   on both actions (5) and (\ref{pary2.1}) - the last term in (5)
     shall introduce 
     the
   functional Dirac delta function replacing $v_i$ by $\dot{x}_l$.}

\begin{equation}\label{pary2.5}
  { L}^{(0)} = \frac{mv^2_i}{2} - k {\epsilon}_{ij} v_i
  \dot{v}_j +   p_i (\dot{x}_i - v_i ) \, ,
\end{equation}
with six canonical variables $(x_i, v_i, p_i)$. The canonical
quantization of (5) using Dirac brackets
 leads again to the relations (4).

Next we introduce the variables [5,6]
\begin{eqnarray}\label{pary2.6}
  Q_i &= & - 2k(v_i - p_i) \, ,  \qquad P_i = p_i
  \cr
  X_i &=  & x_i + \epsilon_{ij} Q_j \, ,
\end{eqnarray}
one gets the following set of canonical Poisson brackets (PB)
\begin{eqnarray}\label{pary2.7}
\left\{  X_i, X_j \right\} &= &  -2k \epsilon_{ij}\, ,
\qquad \qquad
  \left\{  P_i, P_j \right\}  =  0\, ,
  \cr
  \left\{  X_i, P_j \right\} &= &
  \delta_{ij}\, ,
\end{eqnarray}
and
\begin{equation}\label{pary2.8}
\left\{  Q_i, Q_j \right\} =  2k \, \epsilon_{ij}\, .
\end{equation}
The action (5) takes the form
\begin{equation}\label{pary2.9}
  L^{(0)}=  L^{(0)}_{\rm ext}  +       L^{(0)}_{\rm int}\, ,
\end{equation}
where
\begin{eqnarray}\label{pary2.10}
  L^{(0)}_{\rm ext}  &= & P_i \dot{X}_i 
   - k  \epsilon_{ij}
   \, P_i \dot{P}_j -
   \frac{1}{2} \overrightarrow{P}^2\, ,
   \\
      L^{(0)}_{\rm int}  &= & 
      - \frac{1}{4k}\, \epsilon_{ij} Q_i
\dot{Q}_j +
   \frac{1}{8k^2} \overrightarrow{Q}^2\, .
   \label{pary2.11}
\end{eqnarray}
We note that the external and internal degrees of freedom are
dynamically independent, and following Duval and Horvathy [7] we 
can consider  the part (10) of the action as the first order
action describing noncommutative particles. We observe that our 
 model permits easily the consistent  introduction of a scalar
 potential [5], electromagnetic interactions  [5--7] and general
 Lagrangian framework [8].

We see that the action (10) describes an invariant sector of the
model (\ref{pary2.1}) in ``external'' phase space $X_i, P_i$.
  The  internal degrees of freedom (11) can be  related with
 nonvanishing anyonic spin [9]. The aim of this note is to
 supersymmetrize both actions (\ref{pary2.1}) and (10) and discuss 
the relation
between such supersymmetric models. We will show that the split
(9) into dynamically independent supersymmetric
  parts with external and internal
degrees of freedom   can be performed again.

\section{Supersymmetrization of Higher
  Order Action and its First Order Form}

Let us consider for simplicity N=1 supersymmetric quantum
mechanics. We introduce the real field $X_i(t,\theta)$ with one
Grassmann variable $\theta$
\begin{equation}\label{pary2.12}
  x_i (t) \longrightarrow X_i(t, \theta) = x_i (t) + i \theta
   \psi(t)\, ,
\end{equation}
where $\theta^2
 =  \theta \psi_j +
  \psi_j \theta  
 =
   \psi_i \psi_j  + \psi_j \psi_i =0$.
Introducing  the supersymmetric covariant derivative
\begin{equation}\label{pary2.13}
  D = \frac{\partial}{\partial \theta} - i \theta
  \frac{\partial }{\partial t}
  \Rightarrow D^2 = - i \frac{\partial }{\partial t}= - H\, ,
\end{equation}
we get the following supersymmetric extension of  (\ref{pary2.1})
\begin{eqnarray}\label{pary2.14}
L^{(0)}_{\rm SUSY} &= &i \int d\theta \Big( \frac{m}{2} \dot{X}_i
DX_i - k \epsilon_{ij} \ddot{X}_i DX_j  \Big) \cr &= & \frac{m}{2}
\Big( \dot{x}^2_i + i \dot{\psi}_i \psi_i \Big) - k
\epsilon_{ij}\Big( \dot{x}_i \ddot{x}_j - i \dot{\psi}_i
\dot{\psi}_j \Big)\, .
\end{eqnarray}
If $k=0$ we obtain the standard case of N=1 nonrelativistic spinning
particle, with fermionic second class constraints. If $k\neq 0$
the fermionic
 momenta become independent from fermionic
 coordinates $\psi_i$.
 
  Using
   the 
  Faddeev-Jackiw method we extend supersymmetrically the
action (5) as follows:
\begin{eqnarray}\label{pary2.15}
L^{(0)}_{\rm SUSY} &= &
 \frac{m v^2_i}{2}
 - k \epsilon_{ij}v_i \dot{v}_j
+ \frac{im}{2} \psi_i \rho_i
 \cr  &&+ \, i\, k \epsilon_{ij} \rho_i
\rho_j
 + p_i (\dot{x}_i - v_i ) + \chi_i( \dot{\psi}_i - \rho_i)\, .
\end{eqnarray}
The field equation for $\rho_i$ is purely algebraic
    where $\rho_i$ and $\chi_i$ are fermionic.
 Substituting
\begin{equation}\label{pary2.16}
  \chi_i = \frac{im}{2} \psi_i - 2ik\, \epsilon_{ij} \rho_j\, ,
\end{equation}
 we find that (we put for simplicity further $m=1$)
\begin{eqnarray}\label{pary2.17}
L^{(0)}_{\rm SUSY} &= &
 \frac{ v^2_i}{2}
 - k \epsilon_{ij}v_i \dot{v}_j
+ \frac{i}{2} \psi_i \dot{\psi}_i \cr  &&+ \,
2i k \epsilon_{ij} \dot{\psi}_i {\rho}_j
 - i k \epsilon_{ij} \rho_i {\rho}_j
+p_i
 ( \dot{x}_i - v_i)\, .
\end{eqnarray}
 For 
 the fermionic
  sector of the action (17) one obtains
    with the use of Dirac brackets
    the following  PB algebra 
  \begin{equation}\label{pary2.18}
  \{ \psi_i , \psi_j \} = 0 \, , \qquad
   \{ \psi_i, \rho_j \} = \frac{i}{2k}\, \epsilon_{ij} \, ,
   \qquad
   \{\rho_i, \rho_j \} = i \frac{1}{4k^2} \, \delta_{ij}
    \end{equation}
In order to split (15)  into external and internal sector we introduce
 besides the variables (6) also new fermionic variables
\begin{equation}\label{pary2.19}
  \psi_i \longrightarrow\widetilde{\psi}_i = \psi_i 
  -   2k\epsilon_{ij}
  \rho_j \, .
\end{equation}
  We  get
\begin{equation}\label{pary2.20}
L^{(0)}_{\rm SUSY}  = L^{(0)}_{\rm SUSY; ext} + L^{(0)}_{\rm
SUSY;int}\, ,
\end{equation}
where
\begin{equation}\label{pary2.21}
  L^{(0)}_{\rm SUSY;ext} = P_i \dot{X}_i - k
   \epsilon_{ij}P_i\dot{P}_j
 - \frac{1}{2}  P^2_i
   + \frac{i}{2} \widetilde{\psi}_i
  \dot{\widetilde{\psi}}_i \, ,
\end{equation}

\begin{equation}\label{pary2.22}
 L^{(0)}_{\rm SUSY;int} = - \frac{1}{4k}
 \epsilon_{ij}Q_i\dot{Q}_j
 + \frac{1}{8k^2} \overrightarrow{Q}^2 - 2ik^2 \rho_k \dot{\rho}_k
  - ik \epsilon_{ij}\rho_i \rho_k \, ,
\end{equation}
 The new fermionic coordinates satisfy the following PB algebra
\begin{eqnarray}\label{pary2.23}
&  \left\{ \widetilde{\psi}_i, \widetilde{\psi}_j \right\} =
  i\delta_{ij}\, ,
\qquad
  &  \left\{ \widetilde{\psi}_i, {\rho}_j \right\} =
  0 \, .
\end{eqnarray}
We see therefore that again the supersymmetric action (15) or
(20) can be split into dynamically independent external and
internal sectors (see (21)--(22)).

\section{Supersymmetry in External and Internal Sectors}

The actions (21) and (22) describe the supersymmetric extensions
respectively of external and internal actions (10) and (11).
These actions are invariant under the following set of
supersymmetry transformations:

i) in external sector (see (21))
\begin{eqnarray}\label{pary2.24}
\delta\, X_i &= & i\, \epsilon \, \widetilde{\psi}_i \, , \cr
\delta\, \widetilde{\psi}_i &= & - \,  \epsilon \, {P}_i \, , \cr
\delta P_i &=& 0
\end{eqnarray}

ii) in internal sector (see (22))

\begin{eqnarray}\label{pary2.25}
\delta\, Q_i &=&  2i \, k \, \epsilon \, \epsilon_{ij} \, \rho_j
\, ,
\cr
\delta \rho_i &=&  \frac{1}{4k^2}  \, \epsilon \, Q_i\, ,
\end{eqnarray}
where $\epsilon$  is a constant Grassmann number.

The supercharge corresponding to (22) is given by the formula
\begin{equation}\label{pary2.26}
  Q_{\rm ext} = i \widetilde{\psi}_i P_i
\end{equation}
and we get consistently the external Hamiltonian (see (21))

\begin{equation}\label{pary2.27}
-  \frac{i}{2}
\{Q_{\rm ext}, Q_{\rm ext} \}  
 =
  \frac{1}{2} {P}_i^2 = H_{\rm SUSY;ext}^{(0)}\, .
\end{equation}
Similarly from (7) and (22) we obtain the tranformation (25) if
\begin{equation}\label{pary2.28}
Q_{\rm int} = i Q_i \, \rho_i
\end{equation}
and our internal Hamiltonian (see (22)) is given by
\begin{equation}\label{pary2.29}
-  \frac{i}{2}                
\{Q_{\rm int}, Q_{\rm int} \} 
=
- \frac{1}{8k^2} \overrightarrow{Q}^2 
 + i k\, \epsilon_{ij}
\rho_i \rho_j = H_{\rm SUSY;int}^{(0)} \, .
\end{equation}
We would like to add  here that one could consider only the 
external
 part, described by the action (20), as describing supersymmetric 
planar
 particles. The quantum mechanical states describing
  the  internal sector   can be 
 eliminated by  subsidiary conditions.

\section{Final Remarks}
We would like to mention that

i) We have  considered here the N=1 world line supersymmetry. It is
quite straightforward to extend the above considerations to N=2 by
 employing  the N=2 D=1 superfields.

 ii) We have  discussed here, for simplicity, only the free case. The
 supersymmetrization of the models with gauge intersections 
considered
  in [6 ] is under active  consideration.

\end{document}